\documentclass[11pt]{article}
\usepackage{epsfig}

\def\la{\langle}
\def\ra{\rangle}
\def\d{{\rm d}}
\def\dd{\partial}
\def\db{\bar\partial}
\def\t{\theta}
\def\tb{\tilde\theta}
\def\O{{\cal O}}
\def\Z{{\Bbb Z}}
\def\a{\alpha}
\usepackage{oldgerm}
\input{amssym}
\def\be{\begin{equation}}
\def\ee{\end{equation}}
\def\bea{\begin{eqnarray}}
\def\eea{\end{eqnarray}}
\newcommand\egal{&\!\!\!=\!\!\!&}

\newcommand\plus{&\!\!\!+\!\!\!&}

\catcode`\@=11
\def\numberbysection{\@addtoreset{equation}{section}
     \def\theequation{\thesection.\arabic{equation}}}
\numberbysection

\newcommand\Cont[1]{\hsp{.4}\vtop{\ialign{##\crcr$\hfil\displaystyle
                   {\hsp{-.4}#1}\hfil\hsp{-.4}$\crcr
                   \noalign{\kern-1.9pt\nointerlineskip\vskip2pt}
                   \leftrighthookfill\crcr}}\hsp{.4}}
\newcommand\hsp[1] {\mbox{\hspace{#1 em}}}
\def\lefthook      {{\vrule height5pt width0.4pt depth0pt}}
\def\righthook     {{\vrule height5pt width0.4pt depth0pt}}
\def\leftrighthookfill{$\mathsurround=0pt \mathord\lefthook
                   \hrulefill\mathord\righthook$}

\newsavebox{\PSLASH}
\sbox{\PSLASH}{$p$\hspace{-1.8mm}/}

\begin{document}
\title{Abelian Sandpile Model on the Honeycomb Lattice}
\author{N. Azimi-Tafreshi\footnote{e-mail: azimi@physics.sharif.ir}, H. Dashti-Naserabadi,
 S. Moghimi-Araghi,\footnote{e-mail: samanimi@sharif.edu}\\\small{Department of Physics, Sharif University of Technology, Tehran,
P.O.Box: 11365-9161, Iran}
\\\\ P. Ruelle \\
\small{Institut de Physique Th\'eorique, Universit\'e catholique de
Louvain, B-1348 Louvain-La-Neuve, Belgium }}\date{} \maketitle
\begin{abstract}
We check the universality properties of the two-dimensional Abelian sandpile model
 by computing some of its properties on the honeycomb lattice. Exact expressions
 for unit height correlation functions in presence of boundaries and for different boundary
 conditions are derived. Also, we study the statistics of the boundaries of avalanche
  waves by using the theory of SLE  and suggest that these curves are conformally invariant and described by SLE$_{2}$.
\vspace{5mm}%
\end{abstract}
\section{Introduction}
Bak, Tang and Wiesenfeld introduced the theory of self-organized
criticality as a general mechanism that can explain the
behaviour of complex systems which naturally organize themselves into
a critical state \cite{BTW}. They defined the sandpile model as an
example of slowly driven and dissipative complex system to explain
the concept of self-organized criticality. Thanks to the Abelian
property of the model \cite{Dhar}, many statistical and
dynamical results have been derived exactly. Among the main analytical results obtained for the isotropic two-dimensional model, one can
mention 1-site probabilities of height variables \cite{priez}, bulk
correlations of height 1 variables and of some specific clusters known
as weakly allowed clusters \cite{correlation,ASMc2,jeng1}, the discussion of boundary conditions \cite{BC,wind} and
 the effect of boundaries on height probabilities \cite{BC2,BC3,jeng3,conformalb,jpr}, boundary
correlations of height variables \cite{boundCorre,piru,jeng2}, bulk correlations of higher height variables \cite{corre} and
avalanche and toppling wave distributions
\cite{wave1,avalanche1,avalanche2,avalanche3}. The sandpile model has
been investigated in the continuum limit and from a field
theoretical perspective. It has been shown that the Abelian sandpile
model can be described by a specific logarithmic conformal field theory, with central charge $c=-2$ \cite{ASMc2,actionASM,jpr,corre}.

Most of the results, analytical or numerical, have been obtained when the sandpile model is formulated on the square lattice. In fact, the
simple and symmetric structure of the square lattice make it easier to
carry out some of the lattice calculations. However, other regular two-dimensional lattices have been considered.
Height probabilities and critical exponents of the sandpile model on
the triangular and honeycomb lattices have been investigated using
renormalization group transformations and numerical simulations
\cite{RG,papo}. Moreover the critical exponents and the finite-size
scaling functions for the avalanche wave distributions on the
square, honeycomb, triangular, and random lattices have
 been evaluated from Monte Carlo simulations \cite{universality}.
The results clearly suggest that the model on different lattices has
the same set of critical exponents and scaling behaviour.

In this paper, we study the Abelian sandpile model on the honeycomb
lattice. The structure of the honeycomb lattice is more complicated than the
square lattice, as there are two lattice points in each unit cell. However, the correspondence between
 recurrent configurations and spanning trees is maintained, and
an exact expression for the Green function can be obtained, along with the exact asymptotic value.
 These are the two main ingredients to perform the lattice calculations and in turn, to check explicitly
 the universal behaviour of the model. We do that by looking
at the universal terms of the unit height correlation functions with and without boundaries.

The universality in the critical behaviour of sandpile model can be
also checked from the point of view of geometrical features of the
model. Indeed the dynamics of the model is such that each avalanche
is formed on a compact domain with a boundary which converges to a
fractal curve in the scaling limit \cite{compact}. For the sandpile
model defined on the square lattice, it has been recently suggested
that the boundary of these avalanche clusters belongs to the family
of conformally invariant curves generated by the Schramm-Loewner
evolution process SLE$_{\kappa}$, for a diffusitivity constant
$\kappa=2$ \cite{sandpileSLE}.

Avalanches can be also decomposed into a sequence of simpler objects
called toppling waves. While avalanches are believed to be described
by a multifractal set of scaling exponents and have a complex
scaling behaviour, waves show simple scaling properties and are more
convenient for the analysis of avalanche statistics.
 Here, we also check the universality properties by studying the statistics of
  toppling wave boundaries for the model defined on the honeycomb lattice.

This paper is organized as follows: in the next section we define the
sandpile model on the honeycomb lattice and give the exact (and well-known) expression
 for the discrete Green function on the lattice. In the third section, we review the
  methods used for our lattice calculations and then give the details of the results in Section $4$.
We compare our results with the predictions of the $c=-2$ conformal
field theory in Section $5$. Finally, we investigate in Section 6
the statistics of the toppling wave boundaries on the honeycomb
lattice and verify its universality. We finish with some
conclusions. The values of the lattice Green function for small
distances are listed in an Appendix, which also contains the details
for the calculations of its asymptotic behaviour for large
distances.

\section{The sandpile model}

We start by defining the sandpile model on a two-dimensional honeycomb lattice of linear
 size $L$ and with $N$ sites. To each site of the lattice a
random integer variable $h\in \{1,2,3\}$ is assigned which can be
interpreted as the number of sand grains at that site. A
configuration is characterized by the set of heights of all sites and
is called stable if the height values are equal to
1, 2 or 3.

The dynamics of the model is defined in two steps:
$1)$ given a stable configuration, a grain of sand is added to a
randomly chosen site, so that its height is increased by one, while the other sites remain
 unchanged; $2)$ if the height of that site
becomes greater than the critical height $h_{c}=3$, the site becomes
unstable, topples and loses three grains of sand, each one of which drops
on one of the nearest neighbours. The toppling rule can be written in the
form $h_{j} \to h_{j}-\Delta_{ij}$ for all sites $j$, where
$\Delta_{ij}$ is called the toppling matrix and is equal to the
discrete Laplacian, namely
\be
\Delta_{ij} = \cases{
3 & if $i=j$,\cr
-1 & if $i$ and $j$ are nearest neighbours,\cr
0 & otherwise.}
\label{lapl}
\ee
If, as a result of this toppling, some of the
neighbours become unstable, the toppling process continues until all
sites become stable and the avalanche ends. Thus in each time step,
the dynamics takes the system from a stable configuration to another
stable configuration.

For a lattice with $N$ sites, the total number of stable
configurations is $3^{N}$, but not all of them occur in the steady
state. The stable configurations are divided into two classes:
recurrent and transient. After a long time, when the system enters
the steady state, the transient configurations have zero
probability of occurrence and all recurrent configurations occur
with equal probability \cite{Dhar}. The burning algorithm allows to determine whether
 a given configuration is recurrent or not, and also establishes
a one-to-one correspondence between recurrent states and spanning
trees. Thus one can compute the total number of recurrent
configurations by enumerating the spanning trees. From Kirchhoff's theorem, the number
 of spanning trees is given by the determinant of the toppling matrix $\Delta$, or the discrete Laplacian matrix.

If we keep the form of the toppling matrix as above in (\ref{lapl}),
even for the boundary sites which have strictly less than three
neighbours, then sand grains will leave the system each time a
boundary site topples. Such sites are therefore dissipative and
called open. If we set the diagonal element $\Delta_{ii}$ equal to
the number of  neighbours of $i$, then $i$ is conservative and
called closed. Boundary sites can freely be chosen open or closed;
the dynamics of the model, described above, is well-defined provided
the system contains at least one open site.


\begin{figure}[t]
\centering
\begin{tabular}{cc}
\epsfig{file=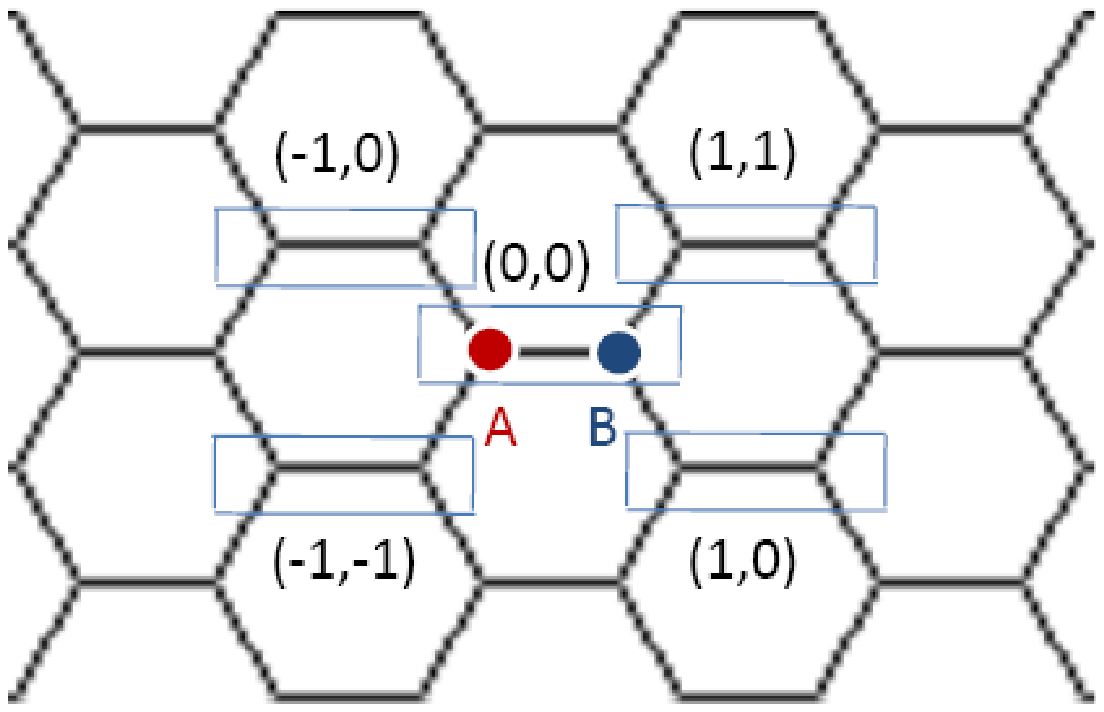,width=0.4\linewidth} &
\epsfig{file=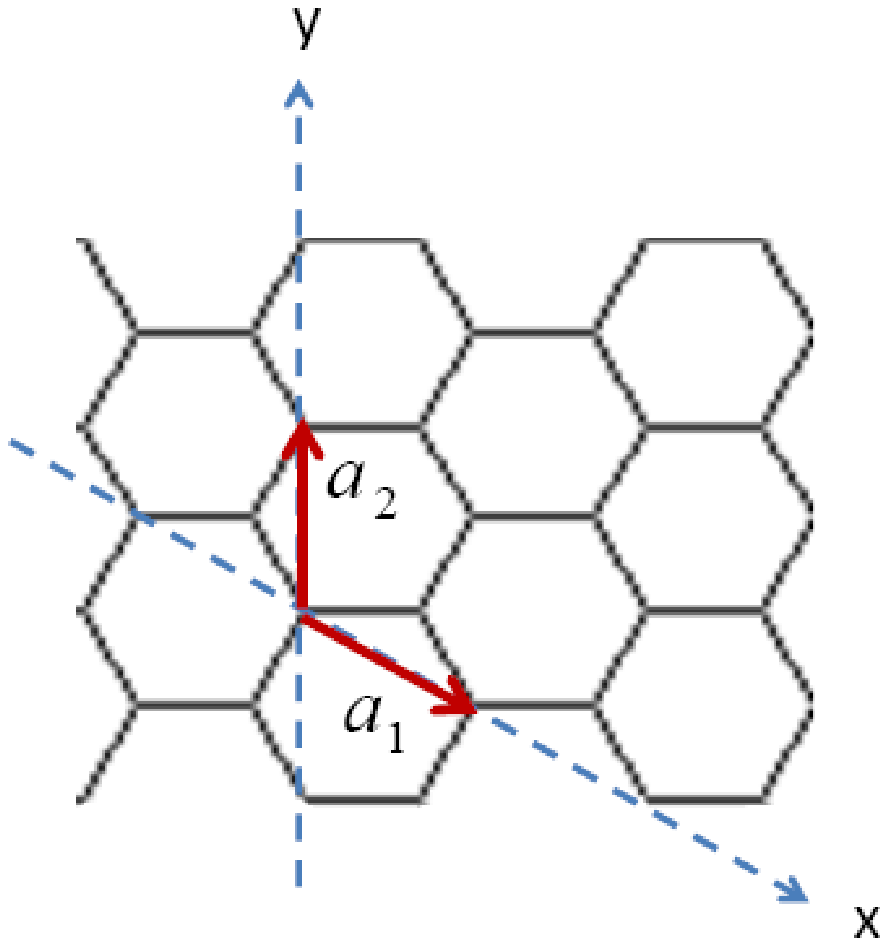,width=0.35\linewidth}\\
$(a)$ & $(b)$ \\
\end{tabular}
\caption{$(a)$ A portion of the honeycomb lattice with the unit
cells, marked by rectangles. Each unit cell contains two types of
lattice sites, $A$ and $B$. $(b)$ The coordinate system used to
label the cells is spanned by the two unit vectors $\vec a_{1}$ and
$\vec a_{2}$.}
\end{figure}

The honeycomb lattice can be divided into unit cells such that each
cell contains two lattice points, one of type $A$ (left site) and
one of type $B$ (right site), as shown in Fig.1a. We choose the
origin of the coordinate system at a site of type $A$. The position
of a cell is specified by the vector $n\vec a_1 + m\vec a_2$, where
$\vec a_{1}$ and $\vec a_{2}$ are the unit vectors shown in Fig.1b
and $n,m$ are integer numbers in $\Z$. Then, each site is
characterized by the location of its cell, $\vec r=(n,m)$, and its
location inside the cell, as $\alpha=0,1$ for $A$-type or $B$-type
sites respectively. So the complete coordinates of a lattice site is
a triplet $(n,m;\alpha)$. We will also use polar coordinates $(n,m)
\leftrightarrow (r,\varphi)$ such that $\vec r = r e^{i\varphi}$,
with $r$ the Euclidean distance to the origin given by $r^2 = n^2 +
m^2 - nm$ and $\varphi$ the angle measured counterclockwise from the
$\hat n$ axis. Explicitly the change of coordinates reads $n=r
\cos{\varphi} + {r \over \sqrt{3}} \sin{\varphi}$ and $m={2r \over
\sqrt{3}}\sin{\varphi}$ (so $(r,\varphi)$ are the polar coordinates
of the site $(n,m;0)$).

The diagonalization of $\Delta$ can be obtained by a method based on
the decomposition of the lattice into unit cells \cite{shrock}. One
can write the connectivity between the vertices of the unit cells
$\vec r_{1}=(n_{1},m_{1})$ and $\vec r_{2}=(n_{2},m_{2})$ in terms
of  $2\times 2$ adjacency matrices $a(\vec r_1;\vec r_2)$ given by
\bea \label{unit} a_{\alpha_1\alpha_2}(\vec r_1;\vec r_2)=\cases{ 1
& if site $\alpha_1$ in cell $\vec r_{1}$ and site $\alpha_2$ in
cell $\vec r_{2}$ are adjacent,\cr 0 & otherwise.} \eea As they only
depend on the difference $\vec r = \vec r_2 - \vec r_1$, we will
write $a(\vec r) \equiv a(\vec r_1;\vec r_2)$. Explicitly the
connectivity matrices of each unit cell with itself and its four
neighboring cells read (rows and columns are labelled by
$\alpha_1,\alpha_2 = 0,1$ in that order) \bea \label{adjacency}
a(0,0)=\pmatrix{0 & 1 \cr 1 & 0}, \qquad a(1,0) = a(1,1) = a^t(-1,0)
= a^t(-1,-1) = \pmatrix{0 & 0\cr 1 & 0},
\end{eqnarray}
while all the other matrices equal to 0. The Laplacian can then be written as
\bea
\label{tensor}
\Delta_{(\vec r_1;\a_1),(\vec r_2;\a_2)} \egal [3 - a(0,0)] \otimes \delta_{\vec r_2,\vec r_1}
- a(1,0) \otimes \delta_{\vec r_2,\vec r_1+\vec a_1} - a(-1,0) \otimes \delta_{\vec r_2,\vec r_1-\vec a_1}\nonumber\\
\noalign{\smallskip}
&& \hspace{.5cm} - \; a(1,1) \otimes \delta_{\vec r_2,\vec r_1+\vec a_1+\vec a_2} - a(-1,-1) \otimes \delta_{\vec r_2,\vec r_1-\vec a_1-\vec a_2}.
\eea

Let us consider an $N_{1}\times N_{2}$ array of unit cells with
periodic boundary conditions, namely the set of cells $\vec r =
(n,m)$ with $1 \leq n \leq N_1$, $1 \leq m \leq N_2$ and periodicity
in both coordinates. In the decomposition (\ref{tensor}), the
matrices acting in the $\vec r$ space are made of cyclic matrices
and are simultaneously diagonalized by going to the basis of
eigenfunctions $v_{k_1,k_2}(n,m) = e^{2i\pi k_1n/N_1} \, e^{2i\pi
k_2m/N_2}$. In this basis, the Laplacian is block-diagonalized,
$\Delta \sim \oplus_{\theta_1,\theta_2} \: M(\theta_1,\theta_2)$,
with \bea M(\theta_1,\theta_2) \egal 3\,{\Bbb I} - a(0,0) - a(1,0)
e^{i\theta_1}
- a(-1,0) e^{-i\theta_1} - a(1,1) e^{i(\theta_1+\theta_2)} - a(-1,-1) e^{-i(\theta_1+\theta_2)}\nonumber\\
\noalign{\medskip}
\egal \pmatrix{3 & -1 - e^{-i\theta_1} - e^{-i(\theta_1+\theta_2)} \cr
-1 - e^{i\theta_1} - e^{i(\theta_1+\theta_2)} & 3},
\eea
where the angles are given by $\theta_j = {2\pi k_j \over N_j}$, with $0 \leq k_1 \leq N_1-1$ and $0 \leq k_2 \leq N_2-1$.

From the previous results, we readily obtain \be \det \Delta =
\prod_{\t_1,\t_2} \det M(\t_1,\t_2) = \prod_{\t_1,\t_2} \Big[6 -
2\cos\t_1 - 2\cos\t_2 - 2\cos{(\t_1+\t_2)}\Big], \ee a number
clearly equal to zero since the eigenvector $v_{0,0}(n,m)=1$ is a
zero mode. If however we leave out the zero eigenvalue, the
Matrix-Tree theorem \cite{bona} states that the product over the
non-zero eigenvalues, divided by the number of sites ($=2N_1N_2$),
yields the total number of spanning trees on the finite array of
cells with exactly one site open. Since $M(0,0)$ has the eigenvalues
$0$ and $6$, we obtain that this number equals
\be {3 \over
N_1N_2}\prod_{(k_1,k_2) \neq (0,0)} \Big[6 - 2\cos{2\pi k_1 \over
N_1} - 2\cos{2\pi k_2 \over N_2} - 2\cos{\Big({2\pi k_1 \over
N_1}+{2\pi k_2 \over N_2}\Big)}\Big].
\ee
From this, one obtains the
entropy per site for the sandpile model on the honeycomb lattice in
the thermodynamic limit \cite{shrock}, \be z =
\int\!\!\!\!\int_{-\pi}^{\pi} \; {\d \t_1 \,\d \t_2 \over 8\pi^2}
\log\Big[6 - 2\cos\t_1 - 2\cos\t_2 - 2\cos{(\t_1+\t_2)}\Big] \simeq
0.807665. \ee The effective number of degrees of freedom per site in
the set of recurrent configurations is thus equal to $e^z \simeq
2.243$.

On the infinite lattice, the Laplacian (\ref{tensor}) depends on
$\vec r = \vec r_2 - \vec r_1$ only, $\Delta_{\a_1,\a_2}(\vec r)
\equiv \Delta_{(\vec r_1;\a_1),(\vec r_2;\a_2)}$, and reads in
Fourier space, \bea \label{Laplacian1}
\widehat{\Delta}_{\a_1,\a_2}(\vec \Theta) \equiv \sum_{\vec r}
\Delta_{\a_1,\a_2}(\vec r) \: e^{i \vec \Theta \cdot \vec r} =
3\,{\Bbb I}-\sum_{\vec r} a_{\a_1,\a_2}(\vec r) \: e^{i \vec \Theta
\cdot \vec r} = M_{\a_1,\a_2}(\t_1,\t_2), \eea where now
$\Theta=(\theta_{1},\theta_{2})$ belongs to $[0,2\pi]^2$. The Green
function is thus the inverse Fourier transform of the inverse of $M$
and also depends on $\vec r$ only, \bea \label{greenhoneycomb}
G_{\a_1,\a_2}(\vec r) \egal
\int\!\!\!\!\int_{-\pi}^{\pi} \; {\d \t_1 \,\d \t_2 \over 4\pi^2} e^{-i \vec \Theta \cdot \vec r} \: M_{\a_1,\a_2}^{-1}(\theta_{1},\theta_{2})\nonumber\\
\noalign{\medskip}
\egal \int\!\!\!\!\int_{-\pi}^{\pi} \; {\d \t_1 \,\d \t_2 \over 4\pi^2}
{e^{-i(n_2-n_1) \theta_{1}}e^{-i(m_2-m_1)\theta_{2}} \over 6 - 2\cos\t_1 - 2\cos\t_2 - 2\cos{(\t_1+\t_2)}}\nonumber\\
\noalign{\medskip}
&& \hspace{3cm} \times \pmatrix{3 & 1 + e^{-i\theta_1} + e^{-i(\theta_1+\theta_2)} \cr
1 + e^{i\theta_1} + e^{i(\theta_1+\theta_2)} & 3}.
\eea
>From this it immediately follows that $G_{AA}(\vec r) = G_{BB}(\vec r)$
and $G_{AB}(\vec r) = G_{BA}(-\vec r)$, as well as
\bea
\label{green1}
G_{AA}(n,m) \egal {3 \over 8\pi^2} \int\!\!\!\!\int_{-\pi}^{\pi} \; \d \t_1 \,\d \t_2  \: {e^{in\t_1 + im\t_2} \over 3 - \cos{\t_1} - \cos{\t_2} - \cos{(\t_1 + \t_2)}},\\
G_{AB}(n,m) \egal \frac{1}{3}[G_{AA}(n,m)+G_{AA}(n+1,m)+G_{AA}(n+1,m+1)],\\
G_{BA}(n,m) \egal \frac{1}{3}[G_{AA}(n,m)+G_{AA}(n-1,m)+G_{AA}(n-1,m-1)],
\end{eqnarray}
in agreement with Poisson's equation.

For the lattice calculations developed in the next sections, we need
to know the Green function for small distances, and its asymptotic
behaviour for large distances. These are discussed and collected in
the Appendix.

\section{Lattice calculations : general formalism}

Our main purpose in this paper is to check the universality
properties of the sandpile model by computing some of its properties
on the honeycomb lattice, for instance joint probabilities of local
height variables. Among these the simplest are related to so-called
weakly allowed clusters (WACs). These are finite clusters of sites
with specific height values such that decreasing the height of any
of their sites by one turns them into forbidden subconfigurations
\cite{Dhar,correlation}.

The method used to compute the joint probability of such clusters,
due to Majumdar and Dhar, is based on a modification of the toppling
matrix \cite{correlation}. They showed that the number of recurrent
configurations which contain a given WAC is equal to the total
number of recurrent configurations of a new sandpile model, defined
in terms of a modified toppling matrix. The modification is usually
done by removing some connections from the cluster to the rest of
the lattice, and at the same time, by adjusting the diagonal
elements of the toppling matrix in order to prevent the sites from
being dissipative. Then the new toppling matrix can be written in
terms of the original one as $\Delta^{\rm new} = \Delta + B$, where
the defect matrix $B$ encodes the modification: $B_{ij} = B_{ji} =
1$ if the symmetric bond between sites $i$ and $j$ is removed, and
$B_{ii}=-n$, if $n$ bonds have been cut from the site $i$. Since the
modifications concern a finite collection of sites, the matrix $B$
has finite rank. The probability of occurrence of a weakly allowed
cluster $S$ is obtained by computing the determinant
\cite{correlation}
\begin{eqnarray}
\label{probability}
P(S)= \frac{\det \Delta^{\rm new}}{\det \Delta} = \det({\Bbb I} + G B),
\end{eqnarray}
where $G = \Delta^{-1}$ is the lattice Green function. Because $B$
has finite rank, the matrix indices in the determinant may be
reduced to the sites affected by the modification, so that the
determinant is actually finite.

The same method can be used to calculate the probability of
occurrence of several WACs, and therefore their correlation
functions. Each cluster $S_i$ comes with its own defect matrix
$B_i$, and the full defect matrix is simply the direct sum of all
$B_i$'s. The matrices $B$ and $G$ acquire a block structure, where
each block refers to the sites involved in each of the clusters. For
example, a $2$-cluster correlation function can be found by
computing \be \label{probtwo} P(S_{1},S_{2})=\det\left({\Bbb I} +
\pmatrix{G_{11} & G_{12} \cr G_{21} & G_{22}} \pmatrix{B_1 & 0 \cr 0
& B_2} \right). \ee If $S_1$ and $S_2$ are respectively located
around $\vec r_1$ and $\vec r_2$, this probability will depend on
Green function entries for small distances within $S_1$ and $S_2$
(those entries in $G_{11}$ and $G_{22}$), and on entries labelled by
pairs of sites, one being close to $\vec r_1$, the other being close
to $\vec r_2$. As one is mainly interested in correlations of
clusters separated by large distances, the calculation of the
determinant requires to know the Green function for both small and
large distances. Calculations of multicluster probabilities, for
about a dozen of different WACs and for up to four clusters, have
been carried out on the square lattice \cite{ASMc2}.

Note that the above formalism remains valid for a finite or infinite
grid, with or without boundaries. The form of $B$ will in general
depend on whether some of the clusters touch a boundary; in addition
the Green function used in the calculations must be appropriate to
the geometry and the boundary conditions used. The thermodynamic
limit can be conveniently evaluated by using the Green function on
the infinite lattice directly.

The simplest of all WACs consists of a single site with height equal
to 1. In  this case, the modification consists in leaving only one
bond between the height 1 and the rest of the lattice, and
correspondingly by decreasing the diagonal entries of $\Delta$ as
explained above. On the honeycomb lattice, it means that $B$ is
identically zero except for a 3-by-3 block \be \pmatrix{-2 & 1 & 1
\cr 1 & -1 & 0 \cr 1 & 0 & -1}, \label{defect1} \ee labelled by the
site where the height is fixed to 1, and any two of its nearest
neighbours. If the site where the height is 1 is on a boundary, the
corresponding non-zero block is smaller.

>From correlations of the above kind, one may infer the scaling behaviour
of lattice observables like the height 1 variable in the bulk or on
a boundary with a given boundary condition. In the scaling limit,
the correlations in the bulk, on a boundary, or at a finite but
large distance from a boundary, are all universal quantities, which
the underlying conformal field theory is able to describe. Being
universal, they should not depend on the type of lattice on which
the model is defined. This is what we want to check in the
following.

On the square lattice, it has been shown \cite{correlation} that the
height 1 variable in the bulk scales like a dimension 2 field,
implying in particular that the 2-point correlation decays like
$1/r^4$. The same is true for the height 1 variable on an open or a
closed boundary \cite{boundCorre,piru}. Once the scaling limit of
the height 1 variable has been properly identified with a specific
field of the conformal theory, higher correlation functions are
fixed. On the square lattice, the lattice results confirm well the
predictions of logarithmic conformal field theory \cite{conformalb}.

In contrast, higher height variables in the bulk do not scale the
same way since their correlations involve logarithmic functions of
the distances \cite{conformalb,jpr,corre}. For instance the 2-point
correlation of a height 1 variable and a height variable strictly
bigger than 1 has been shown to decay like $\log r/r^4$
\cite{corre}, whereas that of height variables greater or equal to 2
is conjectured to decay like $\log^2 r/r^4$ \cite{jpr}. On an open
boundary, all four height variables scale the same way, while on a
closed boundary, the height 2 and 3 variables have a slightly
different but still non-logarithmic scaling compared to the height
1. All known results on the square lattice are consistent with the
identification of the height 2, 3 or 4 variables in the bulk as the
logarithmic partner of the height 1.

Likewise on the honeycomb lattice, height 2 and 3 variables are
expected to have a logarithmic scaling similar to the heights 2, 3
and 4 on the square lattice. Here we will restrict ourselves to the
height 1, and check that it can be identified with the same
conformal fields as on the square lattice.




\section{Lattice calculations : results}

The calculation of height $1$ probability is the simplest case. We
give in this Section the results we have obtained for the various
correlations, in the bulk with and without boundaries, and along
boundaries.

\subsection{In the bulk}

As recalled above, if we want to have a height $1$ at a site $i$, we
should remove the bonds between $i$ and two of its neighbours, and
change the diagonal entries of the toppling matrix. This corresponds
to take the defect matrix $B$ as above in (\ref{defect1}). This
defect matrix is non-zero on three sites only, namely $i$ and the
two chosen neighbours, so that the general formula requires to know
the Green function on the same three sites. From the results of the
Appendix and obvious symmetries, it reads, for the infinite lattice
(we assume $i$ is of type $A$), \be G = \pmatrix{G_{AA}(0,0) &
G_{AB}(0,0) & G_{AB}(0,0) \cr G_{AB}(0,0) & G_{AA}(0,0) &
G_{AA}(1,0) \cr G_{AB}(0,0) & G_{AA}(1,0) & G_{AA}(0,0)} =
G_{AA}(0,0) + \pmatrix{0 & -{1 \over 3} & -{1 \over 3}\cr -{1 \over
3} & 0 & -{1 \over 2}\cr -{1 \over 3} & -{1 \over 2} & 0}. \ee The
constant piece $G_{AA}(0,0)$ is divergent on the infinite lattice,
but drops out in the product $GB$ since $B$ has column (and row)
sums equal to 0. We thus find that the probability that a given site
has a height equal to 1 is given, in the infinite volume limit, by
\be P_1 = \det({\Bbb I} + GB) = \frac{1}{12} \simeq 0.0833. \ee The
other two single height probabilities are known from numerical
simulations \cite{universality}, and found to be $P_2 \simeq 0.2937$
and $P_3 \simeq 0.623$.

We can similarly compute multi-site height 1 probabilities. We first
consider the two-site probability for two A-type sites, one at the
origin and the other at $(n,m;0)$. The two-site height $1$
probability is obtained according to the formula in Eq.
(\ref{probtwo}), for which the Green function for sites close to the
two reference points is required. This is computed for large
distances and arbitrary positions in the Appendix, and we
obtain\footnote{The analogous calculations on the square lattice
\cite{ASMc2} have been carried for specific spatial configurations
of the heights 1, mostly when they are aligned on a principal or
diagonal axis. The use of the Green function for arbitrary positions
from the Appendix enables us to read off more clearly the universal
terms and allows a more direct comparison with conformal field
formulas.} \be \label{two-pointAA} P_{11}(n,m;0) = P_1^{2}\left(1 -
\frac{3}{2\pi^{2}}\frac{1}{r^{4}}
-\frac{1}{2\pi^{2}}\frac{2+5\cos{6\varphi}}{r^{6}}+\ldots\right).
\ee

The two-site probability when the distant site is of type $B$ can be
computed in a similar way, and reads \be \label{two-pointAB}
P_{11}(n,m;1) = P_1^{2}\left(1 - \frac{3}{2\pi^{2}}\frac{1}{r^{4}}
-\frac{1}{2\pi^{2}}\frac{2 - 5\cos{6\varphi}}{r^{6}}+\ldots\right),
\ee where $r$ and $\varphi$ are the polar coordinates of the site
$(n,m;1)$. We see that the subdominant term $\sim r^{-6}$ not only
depends on the angular position of the distant site but also on its
type, $A$ or $B$. This is a first sign that only the dominant term
in $r^{-4}$ is universal, and rotationally invariant, as expected
for a scalar observable.


In both cases, the 2-point correlation of two heights 1 separated by
a large distance $r$ behaves like \be P_{11}(r) - P_1^2 = -{1 \over
2}\Big({\sqrt{3}P_1 \over \pi}\Big)^2 {1 \over r^4} + \ldots \ee Up
to the numerical coefficient, this is the same result as on the
square lattice: interpreted, in the scaling limit, as the 2-point
correlation function of the field $\phi$ associated to the presence
of a height 1, it implies that $\phi$ is a scalar field with scaling
dimension 2, and fixes its normalization.

The general three-site probability for three heights 1 can also be
computed. Here we restrict ourselves to $A$-sites only, in cells
located at $\vec r_1 = \vec 0, \, \vec r_2$ and $\vec r_3$. The
probability depends on the vectors $\vec r_{ij} = \vec r_i - \vec
r_j$ which we write in polar coordinates as $\vec r_{ij} = r_{ij}
e^{i\varphi_{ij}}$, so that $\varphi_{ij} = \pi + \varphi_{ji}$. We
find the following result for the connected\footnote{The connected
$n$-site probability is obtained by subtracting from the full
$n$-site probability products of lower order probabilities.}
three-site probability, \bea P_{111}(\vec r_1,\vec r_2,\vec
r_3)_{\rm conn} \egal -{1 \over 576\pi^3} {1 \over
(r_{12}r_{13}r_{23})^2} \left\{ {\sin{2(\varphi_{12}-\varphi_{13})}
\cos{3\varphi_{23}} \over r_{23}}\right.
\nonumber\\
\noalign{\medskip}
&& \hspace{0.3cm} + \: {\sin{2(\varphi_{12}-\varphi_{23})} \cos{3\varphi_{13}} \over r_{13}} + \left.{\sin{2(\varphi_{13}-\varphi_{23})} \cos{3\varphi_{12}} \over r_{12}}\right\} + \ldots
\eea
The main observation is the absence of a dimension 6 term, which, in the scaling limit, should correspond to the 3-point function of the field $\phi$. Indeed one sees that on the lattice, the dominant term of the three-site correlation has scale dimension 7 (and even 8 for specific spatial configurations). The same observation was made on the square lattice \cite{ASMc2}, and leads to the expectation that the 3-point function of $\phi$ vanishes identically.

Finally, in the same notations, we have computed the general 4-site probability, again for $A$-sites only. The connected part reads
\bea
P_{1111}(\vec r_1,\vec r_2,\vec r_3,\vec r_4)_{\rm conn} \egal -{1 \over 4}\Big({\sqrt{3}P_1 \over \pi}\Big)^4 \left\{
{\cos{2(\varphi_{13}-\varphi_{14}-\varphi_{23}+\varphi_{24})} \over (r_{13}r_{14}r_{23}r_{24})^2}\right.
\nonumber\\
\noalign{\medskip}
&& \hspace{-2cm} +\: \left. {\cos{2(\varphi_{12}-\varphi_{14}-\varphi_{23}+\varphi_{34})} \over (r_{12}r_{14}r_{23}r_{34})^2} + {\cos{2(\varphi_{12}-\varphi_{13}-\varphi_{24}+\varphi_{34})} \over (r_{12}r_{13}r_{24}r_{34})^2}\right\} + \ldots
\eea
As on the square lattice, the dominant term has the expected scale dimension 8, and should correspond to the 4-point correlator of $\phi$.

\subsection{On upper-half planes}

In addition to probabilities on the infinite lattice, height 1 probabilities on semi-infinite lattices can also be examined. We will consider two upper-half planes, one bordered by a tilted boundary, the other by a horizontal boundary, as shown in Fig.2 and Fig.3. The tilted boundary is parallel to the $\hat n$ axis, and, for this reason, will be called principal; it is made of all $A$-sites on the line $m=1$; each boundary site has two neighbours which lie in the interior of the upper-half lattice. The horizontal boundary contains the sites in all the cells on the line $n=2m-2$; each boundary site has again two neighbours, one of which is itself a boundary site. On each type of boundary, the uniform open or closed condition corresponds to set $\Delta_{ii}=3$ or $\Delta_{ii}=2$ respectively for all boundary sites. Although the technical details differ for the two kinds of boundaries, the results should not, and the joint probabilities should only depend on the distances separating the heights 1 and the boundary.

The calculation of height 1 probabilities on a upper-half plane follows the same principles as on the infinite lattice. In particular the same defect matrix $B$ can be used if the height 1 is not at a boundary site. The only difference is that we should use the discrete Green function adapted to the boundary condition we choose. Open and closed Green functions are usually obtained using the image method, and, except in one case, the same method may be used here too. Let us first consider the principal boundary, in Fig.2.

\begin{figure}[t]
\centering
\begin{tabular}{cc}
\epsfig{file=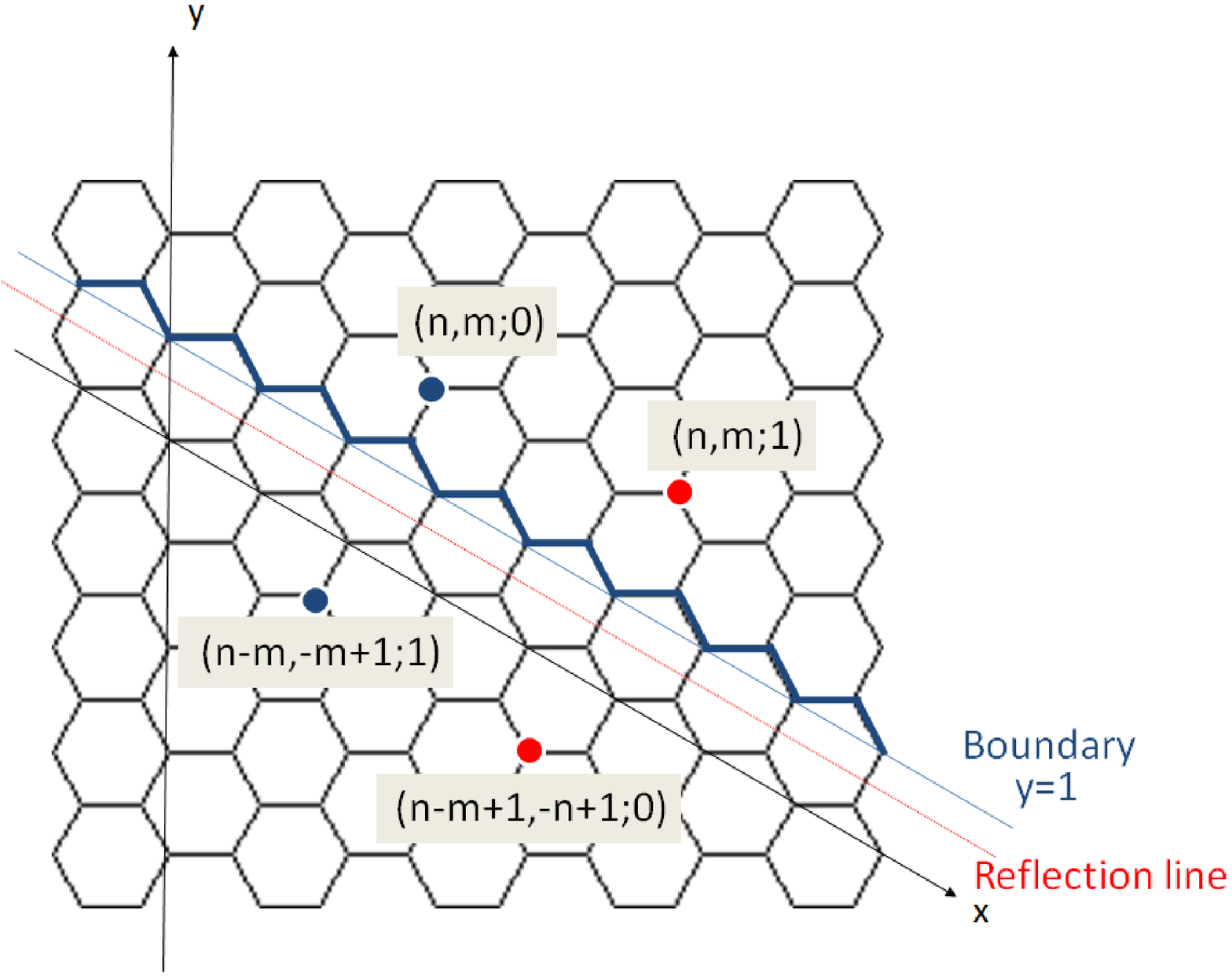,width=0.5\linewidth} &
\epsfig{file=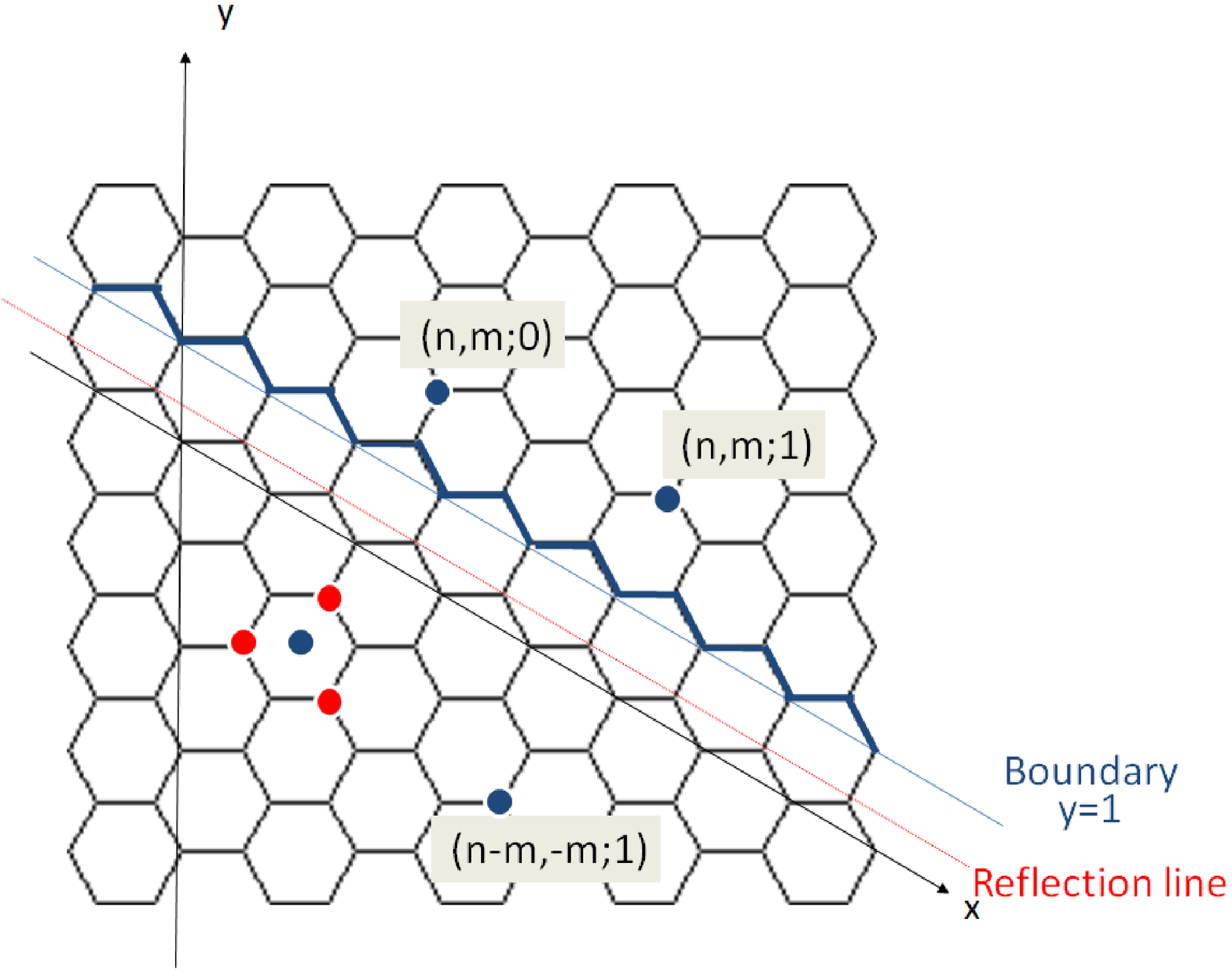,width=0.5\linewidth}  \\
$(a)$& $(b)$ \\
\end{tabular}
\caption{Principal boundary with closed $(a)$ and open $(b)$
boundary condition.}
\end{figure}

For the closed boundary condition, each site of the half-plane above or on the boundary is mirrored through the reflection line shown as the dotted red line. An $A$-site has a mirror image which is a $B$-site and vice-versa. The coordinates of the images are indicated in Fig.2a. The closed Green function for the principal boundary then reads
\bea
\label{openB}
G_{(n_{1},m_{1};\alpha)(n_{2},m_{2};0)}^{\rm cl} \egal G_{(n_{1},m_{1};\alpha)(n_{2},m_{2};0)} + G_{(n_{1},m_{1};\alpha)(n_{2}-m_{2},1-m_{2};1)},\\
\noalign{\smallskip}
G_{(n_{1},m_{1};\alpha)(n_{2},m_{2};1)}^{\rm cl} \egal  G_{(n_{1},m_{1};\alpha)(n_{2},m_{2};1)} + G_{(n_{1},m_{1};\alpha)(n_{2}-m_{2}+1,1-m_{2};0)}.
\eea

For the open boundary condition, the situation is slightly more complicated. The reflection line, shown in Fig.2b as the dotted red line, is such that a $B$-site above the boundary reflects itself to a $B$-site below the boundary, however the mirror image of an $A$-site does not belong to the lattice. Instead the strict mirror of an $A$-site is the center of an hexagon lying below the boundary. We then define the three $B$-sites on this hexagon as the three mirror images of the $A$-site above the boundary, each one being weighted by a factor $1/3$. The open Green function is then related to the Green function on the plane by the following expressions,
\bea
\label{openA}
G_{(n_{1},m_{1};\alpha)(n_{2},m_{2};0)}^{\rm op} \egal G_{(n_{1},m_{1};\alpha)(n_{2},m_{2};0)} - \frac{1}{3}\Big[G_{(n_{1},m_{1};\alpha)(n_{2}-m_{2},-m_{2};1)} + \nonumber\\
&& \hspace{1cm} +\:  G_{(n_{1},m_{1};\alpha)(n_{2}-m_{2}-1,-m_{2};1)}
+ G_{(n_{1},m_{1};\alpha)(n_{2}-m_{2},-m_{2}+1;1)}\Big],\\
\label{openB}
G_{(n_{1},m_{1};\alpha)(n_{2},m_{2};1)}^{\rm op} \egal G_{(n_{1},m_{1};\alpha)(n_{2},m_{2};1)} - G_{(n_{1},m_{1};\alpha)(n_{2}-m_{2},-m_{2};1)}.
\eea

These formulas and the asymptotic behaviour of the bulk Green function enable us to obtain the asymptotic behaviour of $G^{\rm op}$ and $G^{\rm cl}$, and in turn, the scaling form of the height 1 probabilities. The height $1$ probability at an arbitrary point $(n,m;\alpha)$ does not depend on $n$, by translational invariance along the $\hat n$ axis. When $m\gg1$ and for $\alpha=0$, we obtain the following 1-site probabilities,
\bea
\label{prob}
P_{1}^{\rm op}(m) - P_1 \egal +\frac{1}{12\sqrt{3}\pi}\frac{1}{m^{2}}+\frac{1} {18\sqrt{3}\pi}\frac{1}{m^{3}} + \ldots = +\frac{\sqrt{3}P_1}{\pi}\frac{1}{4r^{2}} + \ldots,\\
P_{1}^{\rm cl}(m) - P_1 \egal -\frac{1}{12\sqrt{3}\pi}\frac{1}{m^{2}}-\frac{1} {9\sqrt{3}\pi}\frac{1}{m^{3}}+ \ldots = -\frac{\sqrt{3}P_1}{\pi}\frac{1}{4r^{2}} + \ldots
\eea
where $r = \sqrt{3}(m-1)/2$ is the Euclidean distance from the site $(n,m;0)$ to the boundary. The distinctive change of sign between the two types of boundary conditions was also found on the square lattice \cite{BC2,ASMc2}.

\begin{figure}[t]
\centering \epsfig{file=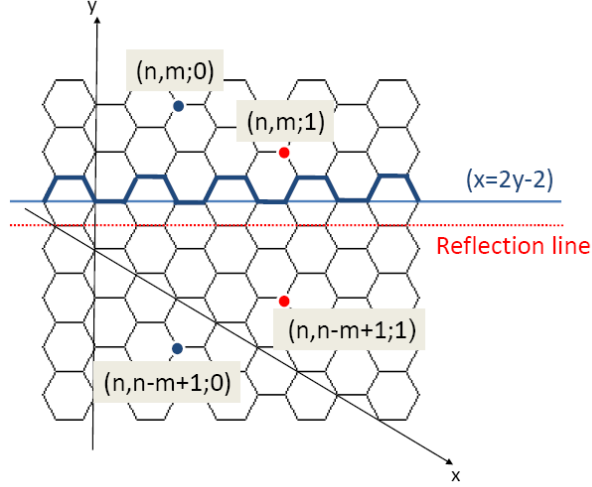,width=0.5\linewidth}
\caption{Horizontal boundary with the open boundary condition.}
\end{figure}

Let us now look at the second, horizontal boundary, shown in Fig.3. We try to adapt the image method to this boundary in order to compute the discrete Green function for open and closed boundary. For the open condition, the image points are found by a reflection with respect to the horizontal line $n=2m-1$, pictured as the dotted red line in Fig.3. Such a reflection preserves the type, $A$ or $B$, of sites, so that we obtain
\be
\label{openBarm}
G_{(n_{1},m_{1};\alpha)(n_{2},m_{2};\beta)}^{\rm op} =
G_{(n_{1},m_{1};\alpha)(n_{2},m_{2};\beta)}-
G_{(n_{1},m_{1};\alpha)(n_{2},n_{2}-m_{2}+1;\beta)}.
\ee

In contrast, for the closed boundary condition, we have not been able to define the appropriate reflection map, with one mirror image or several mirror images like above, so as to make the image method work\footnote{Incidentally, the diagonal boundary on the square lattice, defined by the line $x=y$ in $\Z^2$, may be considered in addition to the more usual boundaries parallel to the principal axes. In this case however, the image method works for both the open and closed boundary conditions. To our knowledge, no calculation has been carried out with this type of boundary.}. As a consequence, we could compute the 1-site height 1 probability for the open condition only. The probability for finding a height $1$ at a site $(2y-2,y+m-1;\alpha)$, located at a distance $r=m-1$ from the horizontal boundary, does not depend on $y$. So we may put $y=1$ and consider the site $(0,m;\alpha)$, in which case the probability reads
\be
\label{poneop}
P_{1}^{\rm op}(m) - P_1 = +\frac{1}{16\sqrt{3}\pi}\frac{1}{m^{2}} + \ldots = +\frac{\sqrt{3}P_1}{\pi}\frac{1}{4r^{2}} + \dots
\ee
with the same dominant term as for the other boundary, as expected.

\subsection{On boundaries}

Finally the height $1$ correlation functions along a boundary can be computed for the types of boundaries considered above. In each case, a boundary site has only two neighbours so that the defect matrix required to force a height 1 is two-by-two and depends on the boundary condition, open or closed. Explicitly they read
\be
B^{\rm op} = \pmatrix{-2 & 1 \cr 1 & -1}, \qquad
B^{\rm cl} = \pmatrix{-1 & 1 \cr 1 & -1}.
\ee

On the principal boundary, it is not difficult, with the appropriate Green functions given in the previous subsection, to compute the one-site probabilities for a height 1, on an open or a closed boundary,
\be
P_1^{\rm op} = \frac{11}{36}+\frac{4}{\sqrt{3}\pi} -\frac{9}{\pi^{2}} \simeq 0.129, \qquad P_1^{\rm cl} = \frac{\sqrt{3}}{\pi}-\frac{1}{3} \simeq 0.218.
\ee
We note that the height of a closed site can only take the values 1 and 2, so that the previous result implies $P_2^{\rm cl} = {4 \over 3} - {\sqrt{3} \over \pi} \simeq 0.782$.

For the joint probabilities of having two, three or four heights 1 separated by distances $r_{ij}$ along the principal boundary, we obtain, for the open boundary condition,
\bea
&& P^{\rm op}_{11}\Big|_{\rm conn} = -{1 \over 4}
\Big({11 \over 2\sqrt{3}\pi}-{9 \over \pi^2}\Big)^2 {1 \over r_{12}^{4}} + \ldots\\
&& P^{\rm op}_{111}\Big|_{\rm conn} = -{1 \over 4}
\Big({11 \over 2\sqrt{3}\pi}-{9 \over \pi^2}\Big)^3 {1 \over (r_{12}r_{13}r_{23})^2} + \ldots\\
&& P^{\rm op}_{1111}\Big|_{\rm conn} = -{1 \over 8}
\Big({11 \over 2\sqrt{3}\pi}-{9 \over \pi^2}\Big)^4 \left\{
{1 \over (r_{12}r_{13}r_{24}r_{34})^2} + {1 \over (r_{12}r_{14}r_{23}r_{34})^2} + {1 \over (r_{13}r_{14}r_{23}r_{24})^2}\right\}\nonumber\\
&& \hspace{8cm} + \ldots
\eea
and similar expressions for the closed boundary condition,
\bea
&& P^{\rm cl}_{11}\Big|_{\rm conn} = -{1 \over 4}
\Big({\sqrt{3} \over 2\pi}\Big)^2 {1 \over r_{12}^{4}} + \ldots\\
&& P^{\rm cl}_{111}\Big|_{\rm conn} = {1 \over 4}
\Big({\sqrt{3} \over 2\pi}\Big)^3 {1 \over (r_{12}r_{13}r_{23})^2} + \ldots\\
&& P^{\rm cl}_{1111}\Big|_{\rm conn} = -{1 \over 8}
\Big({\sqrt{3} \over 2\pi}\Big)^4 \left\{
{1 \over (r_{12}r_{13}r_{24}r_{34})^2} + {1 \over (r_{12}r_{14}r_{23}r_{34})^2} + {1 \over (r_{13}r_{14}r_{23}r_{24})^2}\right\} + \ldots \quad
\eea

On the horizontal boundary, and for the open boundary condition only (as explained above, we cannot handle the closed condition), we obtain similar results for the single site height 1 probability,
\be
P_1^{\rm op} = -\frac{37}{36}+\frac{8}{\sqrt{3}\pi}-\frac{3}{\pi^{2}} \simeq 0.1385,
\ee
and for the first correlation functions,
\bea
&& P^{\rm op}_{11}\Big|_{\rm conn} = -{1 \over 4}
\Big({1 \over \pi^2}-{\sqrt{3} \over 9\pi}\Big)^2 {1 \over r_{12}^{4}} + \ldots\\
&& P^{\rm op}_{111}\Big|_{\rm conn} = -{1 \over 4}
\Big({1 \over \pi^2}-{\sqrt{3} \over 9\pi}\Big)^3 {1 \over (r_{12}r_{13}r_{23})^2} + \ldots\\
&& P^{\rm op}_{1111}\Big|_{\rm conn} = -{1 \over 8}
\Big({1 \over \pi^2}-{\sqrt{3} \over 9\pi}\Big)^4 \left\{
{1 \over (r_{12}r_{13}r_{24}r_{34})^2} + {1 \over (r_{12}r_{14}r_{23}r_{34})^2} + {1 \over (r_{13}r_{14}r_{23}r_{24})^2}\right\}\nonumber\\
&& \hspace{8cm} + \ldots
\eea

\section{Conformal Field Theory}

If the above results for heights 1 on the honeycomb lattice are compared with those obtained on the square lattice \cite{ASMc2,jeng1,piru,jeng2}, it is immediately clear that they coincide, up to normalizations, and thereby confirm the universality of the field assignements. So we restrict here to a brief reminder of the main features of the conformal field interpretation of these lattice results, and take the opportunity to collect the various formulas.

On the square lattice, it has been shown that, in the scaling limit, {\it i.e.} in the large distance limit, the joint height 1 probabilities on the lattice are exactly reproduced by conformal correlators of primary fields. If the height 1 variable lives in the bulk, the corresponding primary field is a non-chiral field with conformal weights $(1,1)$, whereas it is a chiral, boundary primary field of weight 2 in the case the height 1 lies on a boundary, closed or open. It turns out that all these correlators can be understood, and computed, by writing the primary fields in terms of a pair of symplectic free fermions $\t,\tb$.

The theory of symplectic free fermions, with central charge $c=-2$, is the logarithmic conformal field theory which is, by far, the best understood, see for instance \cite{kausch}, and the more recent work \cite{gabrun} as well as the references therein. We only need here the most basic features of it.

The symplectic fermions are anticommuting, space-time scalar fields with propagators given by
\bea
\Cont{\t(z,\bar z) \t}(w,\bar w) \egal \Cont{\tb(z,\bar z) \tb}(w,\bar w) = 0,\\
\Cont{\t(z,\bar z) \tb}(w,\bar w) \egal -\log{|z-w|}, \label{2theta}
\eea
from which all higher correlators may be obtained from Wick's theorem. Since the propagator (\ref{2theta}) is a sum of a chiral term and a antichiral term, the fermions satisfy $\dd\db \t = \dd\db \tb = 0$ in all correlators. A Langrangean realization is provided by the action $S \sim \int \partial\theta\bar{\partial}\tb$, which has the previous two conditions as equations of motion.


The fermion fields themselves are not primary, but their first derivatives are primary. In particular $\phi(z,\bar z) = \dd\db (\t\tb)$ is a primary field with conformal weights $(1,1)$. It is not difficult to compute its 2-, 3- and 4-point functions. They are given explicitly by the following expressions where $z_{ij}=z_i - z_j$,
\bea
&&\la \phi(1) \phi(2) \ra = -{1 \over 2|z_{12}|^4},\\
\noalign{\smallskip}
&&\la \phi(1) \phi(2) \phi(3) \ra = 0,\\
\noalign{\smallskip}
&&\la \phi(1) \phi(2) \phi(3) \phi(4) \ra = {1 \over 4|z_{12}z_{34}|^4} + {1 \over 4|z_{13}z_{24}|^4} + {1 \over 4|z_{14}z_{23}|^4} \nonumber\\
\noalign{\smallskip}
&& \hspace{2.5cm}  -{1 \over 8}\left\{
{1 \over (z_{12}z_{34}\bar z_{13}\bar z_{24})^2} + {1 \over (z_{13}z_{24}\bar z_{14}\bar z_{23})^2} + {1 \over (z_{14}z_{23}\bar z_{12}\bar z_{34})^2} + {\rm c.c.}\right\}.
\eea
The 3-correlator vanishes identically because the various Wick contractions necessarily involve the contraction of $\dd\t$ with $\db\tb$, or $\db\t$ with $\dd\tb$. In the 4-point correlator, the first three terms are products of 2-point functions and are not part of the connected correlator.

The correlation functions of $\phi$ on the upper-half plane can be similarly computed by using the appropriate Green function, namely
\be
\Cont{\t(z,\bar z) \tb}(w,\bar w) = -\log{|z-w|} \pm \log{|z-\bar w|},
\ee
with the $+$ sign for the closed boundary, and the $-$ sign for the open boundary. It yields in particular the 1-point function of $\phi$ on the upper-half plane,
\be
\la \phi(z,\bar z) \ra_{\rm cl \atop op} = \pm {1 \over 4({\rm Im}\,z)^2}.
\ee

A simple comparison with the results obtained in the previous section shows that the leading terms of the connected joint probabilities are exactly reproduced by the above correlators provided the subtracted height 1 variable $\delta(h_i-1) - P_1$ converges, in the scaling limit, to $\alpha \phi(z,\bar z)$ for some normalization $\alpha$. The results in the bulk imply that for the honeycomb lattice, $\alpha = \pm {\sqrt{3} P_1 \over \pi}$. The results on the upper-half planes then fix the sign,
\be
\alpha^{\rm h.c.} = -{\sqrt{3} P_1^{\rm h.c.} \over \pi} = - {1 \over 4\sqrt{3}\pi}.
\ee
By comparison, the results for the square lattice imply $\alpha^{\rm sq} = -P_1^{\rm sq} = -{2(\pi-2) \over \pi^3}$ \cite{ASMc2}. The way this specific conformal field emerges in the scaling limit has been demonstrated in \cite{actionASM}. Moreover, from the conformal field theory point of view, the open and closed boundary conditions have been shown \cite{BC} to be related to each other by the insertion of a chiral primary field of conformal weight $-1/8$, and leads to the change of sign in the 1-point function of $\phi$ on the upper-half plane.

For the purpose of describing the boundary height 1, we need the chiral version of the previous fields. So one also considers chiral symplectic free fermions with contractions
\bea
\Cont{\t(z) \t}(w) \egal \Cont{\tb(z) \tb}(w) = 0,\\
\Cont{\t(z) \tb}(w) \egal -{1 \over 2} \log{(z-w)}.
\eea
The chiral version of $\phi$ that we will use, namely $\phi_{\rm c} = \dd \t \dd \tb$, is not a primary field since it is proportional to the stress-energy tensor of the Lagrangean realization, $T(z) = 2 \, \dd \t \dd \tb$. The first correlators of $\phi_{\rm c}$ with itself read
\bea
&&\la \phi_{\rm c}(1) \phi_{\rm c}(2) \ra = -{1 \over 4z_{12}^4},\\
\noalign{\smallskip}
&&\la \phi_{\rm c}(1) \phi_{\rm c}(2) \phi_{\rm c}(3) \ra = -{1 \over 4} {1 \over (z_{12}z_{13}z_{23})^2},\\
\noalign{\smallskip}
&&\la \phi_{\rm c}(1) \phi_{\rm c}(2) \phi_{\rm c}(3) \phi_{\rm c}(4) \ra = {1 \over 16(z_{12}z_{34})^4} + {1 \over 16(z_{13}z_{24})^4} + {1 \over 16(z_{14}z_{23})^4} \nonumber\\
\noalign{\smallskip}
&& \hspace{2.5cm} -{1 \over 8}\left\{
{1 \over (z_{12}z_{13}z_{24}z_{34})^2} + {1 \over (z_{12}z_{14}z_{23}z_{34})^2} + {1 \over (z_{13}z_{14}z_{23}z_{24})^2}\right\}.
\eea

The boundary 2-, 3- and 4-correlators computed in Section 4.3 have exactly these functional forms, and show that the boundary height 1 variable, subtracted with the appropriate value of $P_1$, converges to $\alpha_{\rm c} \,\phi_{\rm c}$. The normalization depends on the type of boundary, principal or horizontal, and on the boundary condition. One finds
\bea
&&\alpha_{\rm c}^{\rm h.c.,princ,op} = {11 \over 2\sqrt{3}\pi}-{9 \over \pi^2},
\quad\quad
\alpha_{\rm c}^{\rm h.c.,princ,cl} = -{\sqrt{3} \over 2\pi},\\
\noalign{\smallskip}
&&\alpha_{\rm c}^{\rm h.c.,horiz,op} = {1 \over \pi^2}-{\sqrt{3} \over 9\pi}.
\eea
The two normalization factors for the open condition are positive, whereas the normalization for the closed condition is negative.

On the square lattice, the boundary height 1 variable was also seen to converge to $\phi_{\rm c}$ with a normalization, on a boundary parallel to a principal axis, given by \cite{piru,jeng2}
\be
\alpha_{\rm c}^{\rm sq,op} = {6 \over \pi}-{160 \over 3\pi^2}+{1024 \over 9\pi^3}, \quad \quad
\alpha_{\rm c}^{\rm sq,cl} = -{8 \over \pi}\Big({3 \over 4}-{2 \over \pi}\Big).
\ee
Again the normalization is positive for the open, and negative for the closed boundary condition.

It should be emphasized that, whereas the scaling limit of the height 1 variables, in the bulk and on open/closed boundaries, can be described by conformal fields which are themselves related in a simple way to symplectic free fermions, it is not so for all observables of the sandpile model.

On the square lattice, it has been shown that boundary higher height variables scale to conformal fields which have simple expressions in terms of symplectic fermions. On an open boundary, the heights 2, 3 and 4 scale to the same field $\phi_{\rm c}$ as the height 1, while the heights 2 and 3 on a closed boundary have a slightly different scaling, since they converge to a combination of $\phi_{\rm c}$ and $\t \dd\dd \tb$. On the honeycomb lattice, only the field $\phi_c$ is expected (a closed site has two neighbours, and therefore its height takes only two values).

In contrast, the higher height variables in the bulk are all described, up to normalization, by a single scaling field $\psi$, which turns out to be a logarithmic partner of the field $\phi$ describing the height 1 in the bulk. However the reducible but indecomposable representation they generate does not belong to the theory of symplectic fermions\footnote{The recent article \cite{kyri} is a general study, in a much broader context, of classes of (chiral) representations such as the representation generated by the pair $\phi,\psi$, which appears in their Example 7.}. Whether this representation has a Lagrangean realization is an open and important problem. The same distinction between the height 1 and the higher heights in the bulk is expected on the honeycomb lattice, or indeed on any regular lattice.

\section{Boundary of Wave Clusters and Conformal Invariance}


The scaling behaviour of the two-dimensional critical lattice models
can be reflected in the statistics of non-crossing random curves
which form the boundaries of clusters on the lattice. In the 1920's,
Loewner studied simple curves growing from the origin into the
upper-half plane $\Bbb H$ \cite{Lowner}. Loewner's idea was to
describe the evolution of these curves in terms of the evolution of
the analytic function $g_{t}$, which maps conformally the region
outside of the curve into $\Bbb H$. He showed that this function
satisfies the following differential equation
\begin{eqnarray}\label{SLE}
\frac{dg_{t}(z)}{dt}=\frac{2}{g_{t}(z)-\xi_{t}},
\end{eqnarray}
for a real continuous function $\xi_{t}$, related to the image of
the tip of the curve under $g_t$. Conversely, a continuous real
function $\xi_t$ implicitly defines a curve growing in $\Bbb H$.

Much more recently, Schramm followed the idea that a measure on the
continuous driving functions $\xi_t$ would induce a measure on the
set of growing curves in $\Bbb H$, and showed that the latter
measure is conformally invariant if and only if the former measure
is the Wiener measure for the standard one-dimensional Brownian
motion $B_t$ \cite{Schramm}. This subsequently led to a completely
new perspective on random curves arising in conformally invariant
critical systems, see \cite{babe} for a review. In this context,
setting $\xi_{t}=\sqrt{\kappa}B_{t}$ for different parameter
$\kappa$ corresponds to different universality classes of critical
behaviour.

Avalanche boundaries in Abelian sandpile model are random dynamical
curves whose statistics can be studied using the theory of SLE
\cite{sandpileSLE}. It has been suggested, on the basis of numerical simulations on the square lattice, that the boundaries of
avalanche clusters are conformally invariant with the same properties
as loop erased random walk model, with diffusivity constant $\kappa=2$.

Since an avalanche has a complicated structure, understanding its
dynamics can be simplified by decomposing the avalanche into a
sequence of more elementary objects called toppling waves
\cite{wave1}. The waves are constructed as follows. If, as a result
of the addition of a grain to a site $i$, that site $i$ becomes
unstable, it topples, as do the sites which become unstable as a
consequence of the first toppling at $i$. The first wave is the
collection of all sites which have toppled given that the initial
site is not allowed to topple more than once. One can show that the
sites in the first wave all topple exactly once. After the first
wave is completed, the initial site, if still unstable, is allowed
to topple a second time, and doing so, triggers the second wave of
topplings. The process continues, with a third wave, fourth wave and
so on, until the initial site $i$ is stable and the avalanche stops.
The important property of waves is that they are individually
compact (no hole), and the sites in each wave topple exactly once.

\begin{figure}[t]
\centering \epsfig{file=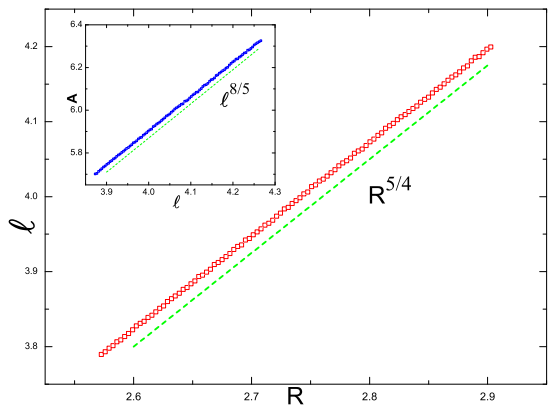,width=0.45\linewidth} \caption{Main
frame: Log-log plot of the length of wave boundaries $l$ versus the
radius of gyration $R$, simulated on the honeycomb lattice with the
linear size of $2048$. Inset: Log-log plot of the average area of
wave clusters $A$ versus the length $l$.}
\end{figure}

To check the universality of the model, we consider an ensemble of
wave boundaries, on the honeycomb lattice, and repeat the analysis
carried out in \cite{sandpileSLE} for the square lattice. At first,
we calculate the fractal dimension $d_{f}$ for the wave boundaries,
determined by the scaling relation $l \sim R^{d_{f}}$ between the
perimeter $l$ of the curve and the radius of gyration $R$. The
result for waves is $d_{f}=1.25\pm 0.01$, see Fig. 4. The inset of
Fig. 4 shows the scaling of the mean area of the wave clusters with
their perimeter length as $A \sim l^{2/d_{f}}$, which is consistent
with the one discussed in \cite{area}. Furthermore, from the
relation $d_{f}=1+\frac{\kappa}{8}$ for the fractal dimension of SLE
curves \cite{SLEdim}, this fractal dimension is consistent with the
value $\kappa=2$, obtained for the boundary of avalanche clusters on
the square lattice \cite{sandpileSLE}. The central charge associated
with $\kappa=2$ is $c={(3\kappa-8)(6-\kappa) \over 2\kappa} = -2$
\cite{CFTSLE}.

One of the questions about SLE curves that has a neat answer is the
following: for a curve connecting two points on the boundary of a
domain $D$, what is the probability that the curve passes to the
left of a given point interior to the domain ? It is usual to take
the domain $D$ to be the upper-half plane and the boundary points to
be the origin and the point at infinity. In this case, an interior
point of the domain is represented in polar coordinates as $z=R
e^{i\phi}$. By scale invariance, the above probability depends only
on $\phi \in [0,\pi]$ and is given by \cite{leftpath}
\begin{eqnarray}
P_{\kappa}(\phi)=\frac{1}{2}+\frac{\Gamma(\frac{4}{\kappa})}{\sqrt{\pi}
\Gamma(\frac{8-\kappa}{2\kappa})}F_{12}\left(\frac{1}{2};\frac{4}{\kappa};\frac{3}{2};-\cot^{2}(\phi)\right)\cot(\phi),
\label{leftpass}
\end{eqnarray}
where $F_{12}$ is the hypergeometric function. For $k=2$, this reduces to
\be
P_2(\phi) = 1 - {2\phi - \sin{2\phi} \over 2\pi}.
\label{left}
\ee

In order to check Eq.(\ref{left}) for wave curves (loop curves), at
first step we should convert these curves to curves which connect
the origin to the infinity (chordal SLE). To this aim, we cross any
given loop by an arbitrary straight line as real line at two points
$x_{0}=0$ and $x_{\infty}$ and consider only a segment of curve
which is above the real line. Then by the conformal map
$\varphi(z)=\frac{x_{\infty}z}{x_{\infty}-z}$, curves in the upper
half plane are transformed to a set of curves connecting the origin
to infinity. \\
The computed probabilities for points at distances $R=0.1,0.2,0.4$
and $0.5$ is consistent with Eq.(\ref{leftpass}), with $\kappa=2.1\pm 0.1$ (see Fig. 5 (a)).\\
\begin{figure}[t]
\centering
\begin{tabular}{cc}
\epsfig{file=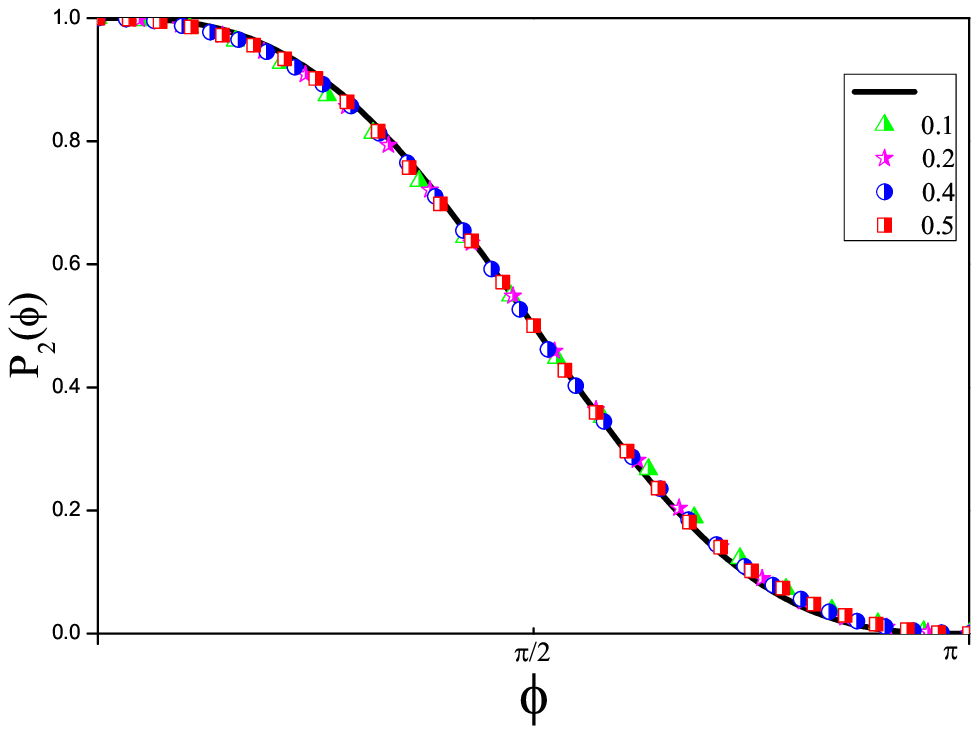,width=0.45\linewidth} &
\epsfig{file=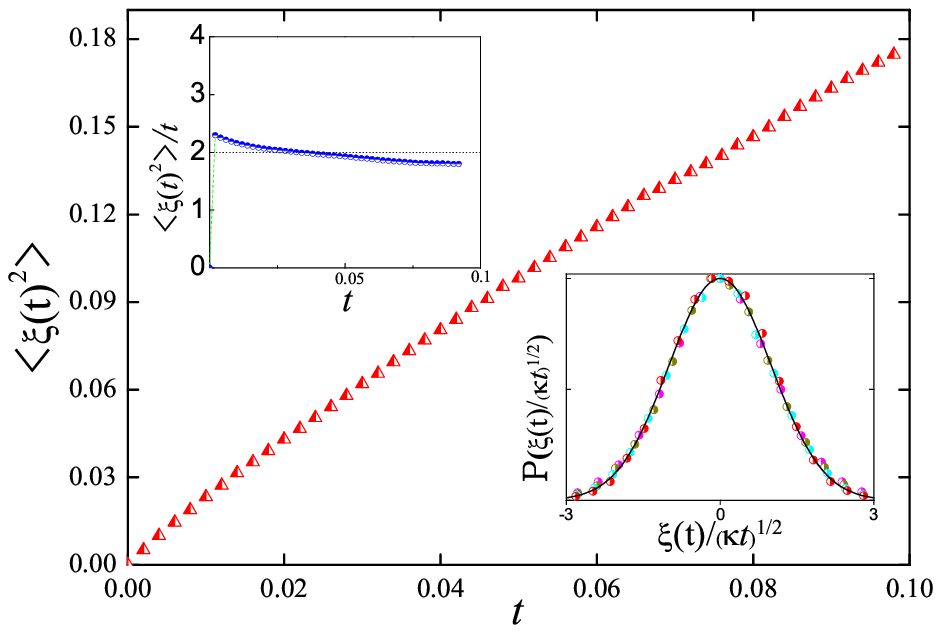,width=0.50\linewidth}  \\
$(a)$& $(b)$ \\
\end{tabular}
\caption{$(a)$ The probability that a wave boundary passes to the
left of a point with polar coordinates $(R,\phi)$, for $R = 0.1,
0.2, 0.4$ and $0.5$. The solid line shows the function $P_2(\phi)$
obtained from SLE for $\kappa = 2$. $(b)$ Statistics of the driving
function $\xi(t)$ for the wave boundaries in the sandpile model on
the honeycomb lattice. Main frame: the linear behaviour of $\langle
\xi(t)^{2}\rangle$ with the slope $\kappa = 2.0\pm 0.2$. Upper-right
inset: the diffusion coefficient is $\kappa=2.0\pm0.2$. Lower-left
inset: the probability distribution function of the noise
$\xi(t)/\sqrt{\kappa t}$, different colors correspond to $t= 0.02,
0.04, 0.06, 0.08.$}
\end{figure}
A more careful test which shows the correspondence with SLE, is to
extract the Loewner driving function $\xi_{t}$. There is an
algorithm for chordal SLE curves, based on the approximation that
driving function is a piecewise constant function \cite{driving}. As
we mentioned at previous case, with conformal map
$\varphi(z)=\frac{x_{\infty}z}{x_{\infty}-z}$, the segment of loop
curves at upper half plane are converted to chordal curves. Then,
each curve is parameterized by the dimensionless parameter $t$ (that
it is not time). In this case the Lowener equation is as
$dg_{t}/dt=2/\{\varphi'(g_t)[\varphi(g_t)-\xi_t]\}$, which
$g_{t}(z)$ maps the half-plane minus the curve up to $t$ into the
whole of upper half-plane. For a constant $\xi$, the equation can be
solved for $g_{t}$ as:
\begin{eqnarray}\label{left2}
G_{t,\xi}(z)= x_{\infty}{\eta x_{\infty} (x_{\infty}-z) +
\sqrt{x_{\infty}^{4}(z-\eta)^{2}+4 t
(x_{\infty}-z)^{2}\times(x_{\infty}-\eta)^{2}} \over
x_{\infty}^{2}(x_{\infty}-z) +
\sqrt{x_{\infty}^{4}(z-\eta)^{2}+4t(x_{\infty}-z)^{2}\times(x_{\infty}-\eta)^{2}}}
\end{eqnarray}
Where, $\eta=\varphi^{-1}(\xi)$.\\
According to algorithm, the interval $[0,T]$ is divided to smaller
intervals $[t_{n},t_{n+1})$ with $t_0=0$ and $t_{N+1}=T$, such that
the $\xi(t)$ is approximated by the constant $\xi_n=\xi(t_n)$ in
each interval. In this case, the function of $g_{t}$ is expressed as
composition of $G_{t_{N}-t_{N-1},\xi_{N-1}}o\ldots o
 G_{t_1,\xi_0}$. The action of each of $G_{t,\xi}(z)$ is such that when
they apply on the points of the curve, remove the first point from
the sequence of points.\\
Now, we follow this algorithm for extracting of the driving
function. At First, we take the points of our curves on upper half
plane approximated with $\{z_{0}=0, z_{1},\ldots, z_{N}=
x_{\infty}\}$. Then, using the parameters
$\eta_{0}=\varphi^{-1}(\xi_{0})=[Re z_{1}x_{\infty}-(Re
z_{1})^{2}-(Im z_{1})^{2}]/(x_{\infty}-Re z_{1})$ and $t_{1}=(Im
z_{1})^{2} x_{\infty}^{4}/\{4[(Re z_{1}-x_{\infty})^{2}+(Im
z_{1})^{2}]^{2}\}$,the map $G_{t_{1},\xi_{0}}$ applied to the points
resulting in a new sequence, by one element shorter:
$z'_{k}=G_{t_{1},\xi_{0}}(z_{k+1})$, with $k=1,\ldots,N$. The
operation is iterated on the new subsequence of points until one
obtains the full set of $t_{k}$ and $\xi_{k}$ for each curve. The
result of this procedure is an ensemble of $\xi(t)$ that its
statistics as shown in Fig. 5 (b) ,converges to a Gaussian process
with variance $\langle \xi^{2}(t)\rangle= \kappa t$ and
$\kappa=2.0\pm 0.2$. This result, together with the other evidences,
certify that the wave boundary curves of sandpile model are
conformally invariant
and described by the SLE$_{2}$.\\
This result seems to be reasonable from the correspondence with the
spanning trees. In fact, one can construct exactly a two-component
tree on the lattice, representing a wave \cite{wave1}. The boundary
of a wave as the dual of the spanning tree is expected to be SLE$_{2}$.
The statistics of the wave boundaries with diffusion coefficient $\kappa=2$
confirm the relation of the sandpile model with a $c = -2$ conformal field
theory.

\section{Conclusions}
In this paper, we have investigated some of the properties of the
Abelian sandpile model on the honeycomb lattice. The scaling
behaviours of the height correlation functions in the bulk, in the
presence of boundaries, and on boundaries, are in the agreement with
those obtained on the square lattice, and correctly predicted by a
$c=-2$ conformal field theory.

We have also checked the universality properties of the model from
the point of view of its geometrical features, namely the statistics
of the boundaries of the toppling waves. We found numerically that
the boundaries of wave clusters are conformally invariant, and well
described by the SLE process with diffusivity $\kappa=2$.

\appendix
\section{Appendix}

We collect in this Appendix some of the values of the lattice Green
function on the honeycomb lattice, for small distances, and also
give its asymptotic behaviour for large distances.

In the coordinate system used in Section 2, the Green function on
the honeycomb lattice for a pair of points of the $A$ type and
separated by the vector $(n,m)$, is given by \be G_{AA}(n,m) = {3
\over 8\pi^2} \int\!\!\!\!\int_{-\pi}^{\pi} \; \d \t_1 \,\d \t_2  \:
{e^{in\t_1 + im\t_2} \over 3 - \cos{\t_1} - \cos{\t_2} - \cos{(\t_1
+ \t_2)}}. \ee One of the two integrations can be carried out, and
leads to the following result \cite{greenasymptotic} \be G_{AA}(n,m)
= {3 \over 2\pi} \int_0^{\pi/2} \; \d x \: {e^{-|n-m|s} \cos{(n+m)x}
\over \sin{x} \, \sqrt{4 - \cos^2{x}}}, \label{GAA} \ee where the
function $s(x)$ is defined through \be \sinh{s} = {\sin{x} \over
\cos{x}} \, \sqrt{4 - \cos^2{x}}. \ee The previous integral is still
divergent, but provides a convergent integral representation for the
subtracted Green function $\Phi_{AA}(n,m) \equiv G_{AA}(n,m) -
G_{AA}(0,0)$. It yields the following values for small $n,m$
\cite{atkin}, \bea
\Phi_{AA}(1,0) \egal \Phi_{AA}(1,1) = -{1 \over 2},\\
\Phi_{AA}(1,2) \egal \Phi_{AA}(-1,1) = 1 - {3\sqrt{3} \over \pi},\\
\Phi_{AA}(2,0) \egal \Phi_{AA}(2,2) = -4 + {6\sqrt{3} \over \pi},\\
\Phi_{AA}(2,3) \egal \Phi_{AA}(-1,2) = {15 \over 2} - {15\sqrt{3} \over \pi},\\
\Phi_{AA}(3,0) \egal \Phi_{AA}(3,3) = -{81 \over 2} + {72\sqrt{3} \over \pi}.
\eea

Let us now evaluate the asymptotic behaviour of $\Phi_{AA}(n,m)$ for
large distances. We will do this calculation by using ideas from
\cite{greenasymptotic} and \cite{jpr}; the analogous calculation for
the square lattice has been done in \cite{gpp}.

The basic idea underlying these computations is that for large
$|n-m|$, the exponential  factor in (\ref{GAA}) contributes
significantly only in the region where $s$ is small, which is also
where $x$ is small. In this region, we may expand $s(x)$ in powers
of $x$, \be s(x) = \sqrt{3} \Big(x + {2 \over 45}x^5 + {2 \over
405}x^9 + \ldots\Big). \ee Therefore the main contribution of the
integral comes from the part close to the origin and is given by the
way the rest of the integrand behaves for small $x$.

We start by splitting the integral giving $\Phi_{AA}(n,m)$ into three pieces,
\bea
\Phi_{AA}(n,m) \egal {3 \over 2\pi} \int_0^{\pi/2} \; \d x \: {e^{-|n-m|s} \cos{(n+m)x - 1} \over \sin{x} \, \sqrt{4 - \cos^2{x}}} \nonumber\\
\noalign{\smallskip}
&& \hspace{-1.5cm} = \; {3 \over 2\pi} \int_0^{\pi/2} \; \d x \: \Big\{{e^{-|n-m|s} \cos{(n+m)x} \over \sin{x} \, \sqrt{4 - \cos^2{x}}} - {e^{-\sqrt{3}|n-m|x} \cos{(n+m)x} \over \sqrt{3} x}\Big\}\nonumber\\
\noalign{\smallskip}
&& \hspace{-1cm} + \; {3 \over 2\pi} \int_0^{\pi/2} \; \d x \: \Big\{{e^{-\sqrt{3}|n-m|x} \cos{(n+m)x} \over \sqrt{3} x} - {1 \over \sqrt{3}x}\Big\}\nonumber\\
\noalign{\smallskip}
&& \hspace{-1cm} + \; {3 \over 2\pi} \int_0^{\pi/2} \; \d x \: \Big\{{1 \over \sqrt{3}x} - {1 \over \sin{x} \, \sqrt{4 - \cos^2{x}}}\Big\}.
\eea

The second integral can be evaluated exactly, up to exponentially small terms, and turns out to give the dominant, logarithmic term, equal to $-{\sqrt{3} \over 2\pi} (\log{r} + \gamma + \log{\pi})$, where $r^2 = n^2 + m^2 - nm$ and $\gamma = 0.577216$ is the Euler constant. The third integral is a constant which can also be computed exactly, and is equal to ${\sqrt{3} \over 4\pi} \log{\pi^2 \over 12}$. We obtain at this stage
\bea
\Phi_{AA}(n,m) \egal -{\sqrt{3} \over 2\pi} \left[\log{r} + \gamma + {1 \over 2} \log{12}\right]
\nonumber\\
\noalign{\smallskip}
\plus {3 \over 2\pi} \int_0^{\pi/2} \; \d x \: \Big\{{e^{-|n-m|s} \cos{(n+m)x} \over \sin{x} \, \sqrt{4 - \cos^2{x}}} - {e^{-\sqrt{3}|n-m|x} \cos{(n+m)x} \over \sqrt{3} x}\Big\}.
\eea

To evaluate the remaining integral, we use the expansion of $s$ as a power series in $x$, and write $e^{-|n-m|s} = e^{-\sqrt{3}|n-m|x} Q(x)$ where $Q$ is expanded as
\be
Q(x) = \exp{[-|n-m|(s - \sqrt{3}x)]} = \exp{\Big\{-\sqrt{3}|n-m|\Big({2 \over 45}x^5 + {2 \over 405}x^9 + \ldots\Big)\Big]}.
\ee
The subtracted Green function then becomes, with $p = |n-m|$ and $q = n+m$,
\bea
\Phi_{AA}(n,m) \egal -{\sqrt{3} \over 2\pi} \left[\log{r} + \gamma + {1 \over 2} \log{12}\right]
\nonumber\\
\noalign{\smallskip}
\plus {3 \over 2\pi} \int_0^{\pi/2} \, \d x \: e^{-\sqrt{3}px} \cos{qx} \, \Big\{{Q(x) \over \sin{x} \, \sqrt{4 - \cos^2{x}}} - {1 \over \sqrt{3} x}\Big\}.
\label{integ}
\eea
By construction, the function in brackets is regular at $x=0$, and may be expanded in powers of $x$. It is not difficult to see from (the polar coordinates have been introduced in Section 2)
\be
\int_0^{\pi/2} \, \d x \: e^{-\sqrt{3}px} \cos{qx} \simeq \int_0^\infty \, \d x \: e^{-\sqrt{3}px} \cos{qx} = {\sqrt{3} p \over 3p^2 + q^2} = {\sqrt{3}\, |\cos{\varphi} - {1 \over \sqrt{3}} \sin{\varphi}| \over 4r},
\ee
where we have neglected exponentially small terms, that the following estimate holds
\be
\int_0^{\pi/2} \, \d x \: x^{k-1} e^{-\sqrt{3}px} \cos{qx} \simeq \O(r^{-k}).
\ee
As a consequence, the integral in (\ref{integ}) has an expansion in inverse powers of $r$, for which the calculation of the $r^{-k}$ terms requires the expansion of the function in brackets to order $k-1$. Because $Q(x)$ has coefficients which depend on $p = \O(r)$, the order $k-1$ means that we keep those terms $p^a x^b$ such that $b-a=k-1$. And since $Q(x)$ is divided by $\sin x \sim x$, $Q(x)$ is to be expanded to order $k$. One easily checks that for fixed $k$, there is only a finite number of terms to consider. The rest of the calculations is straightforward.

To order $r^{-8}$, the relevant expansion of $Q(x)$ reads
\be
Q(x) = 1 - {2p \sqrt{3} \over 45} x^5 - {2p\sqrt{3} \over 405}x^9 + {2p^2 \over 675}x^{10} + \ldots
\ee
from which we obtain the asymptotic expansion of the Green function
\bea
\Phi_{AA}(n,m) \egal -{\sqrt{3} \over 2\pi} \left[\log{r} + \gamma + {1 \over 2} \log{12}\right] \nonumber\\
\noalign{\smallskip}
&& \hspace{2cm} + \: {\sqrt{3} \over 60\pi} \: {\cos{6\varphi} \over r^4} + {5\sqrt{3} \over 168\pi} \: {\cos{6\varphi} \over r^6} + {7\sqrt{3} \over 40\pi} \: {\cos{12\varphi} \over r^8} + \ldots
\eea
We note that it is invariant under the symmetries of the lattice, generated by $\varphi \to \varphi + {2\pi \over 3}$ and $\varphi \to {\pi \over 3} - \varphi$. When $p=0$, corresponding to the line $n=m$ or $\varphi={\pi \over 3}$, the above calculation breaks down. However this line is related by a symmetry of the lattice to $\phi=\pi$ for which $p=|n|$ is not zero. The previous result is therefore valid for all $\varphi$.

As particular cases, we find the asymptotic behaviour on the line $n=m$ ($\varphi = {\pi \over 3}$)
\be
\Phi_{AA}(m,m) = -{\sqrt{3} \over 2\pi} \left[\log{|m|} + \gamma + {1 \over 2} \log{12} - {1 \over 30 m^4} - {5 \over 84 m^6} - {7 \over 20 m^8} + \ldots\right]
\label{diag}
\ee
and on the line $n=2m$ ($\varphi={\pi \over 6}$),
\be
\Phi_{AA}(2m,m) = -{\sqrt{3} \over 2\pi} \left[\log{|\sqrt{3}m|} + \gamma + {1 \over 2} \log{12} + {1 \over 270 m^4} + {5 \over 2268 m^6} - {7 \over 1620 m^8} + \ldots\right]
\ee

For the intended calculations in the sandpile model, we also need to know the Green function for sites in the close neighborhood of a reference site. For this it is sufficient to compute $\Phi_{AA}(n+\ell,m+k)$ with $m,n \gg k,\ell$ as the other entries $\Phi_{AB}$ and $\Phi_{BA}$ may be obtained from them. To compute $\Phi_{AA}(n+\ell,m+k)$, one may simply follow the above calculations in which one appropriately shifts $n$ and $m$ by $\ell$ and $k$ respectively, and then expand the result in inverse powers of $r$. At order 4, we obtain, where $r$ and $\phi$ are the polar coordinates of the site $(n,m)$ as before,
\bea
\Phi_{AA}(n+\ell,m+k) \egal -{\sqrt{3} \over 2\pi} \left[\log{r} + \gamma + {1 \over 2} \log{12}\right] - {\sqrt{3} \over 4\pi} \: {(2\ell-k) \cos{\varphi} + \sqrt{3} k \sin{\varphi} \over r} \nonumber\\
\noalign{\smallskip}
&& \hspace{-3cm} + \: {\sqrt{3} \over 8\pi} \: {(2\ell^2-2k\ell-k^2) \cos{2\varphi} + \sqrt{3} k (2\ell-k) \sin{2\varphi} \over r^2}\nonumber\\
\noalign{\smallskip}
&& \hspace{-3cm} - \:  {\sqrt{3} \over 12\pi} \: {(2\ell^3-3k\ell^2-3k^2\ell+2k^3) \cos{3\varphi} + 3\sqrt{3} k\ell (\ell-k) \sin{3\varphi} \over r^3}\nonumber\\
\noalign{\smallskip}
&& \hspace{-3cm} + \:  {\sqrt{3} \over 240\pi} \: {4\cos{6\varphi} + 15 (2\ell^4-4k\ell^3-6k^2\ell^2+8k^3\ell-k^4) \cos{4\varphi} + 15\sqrt{3} k (4\ell^3-6k\ell^2+k^3) \sin{4\varphi} \over r^4}\nonumber\\
\noalign{\medskip}
&& \hspace{-3cm} + \: \ldots
\eea

\end{document}